 \documentclass[prl,twocolumn,aps,showpacs]{revtex4}

\usepackage{epstopdf}
\newcommand{\3}{$^3$He}

\usepackage{bm}
\usepackage{amssymb}
\usepackage{amsfonts}
\usepackage{amsmath}
\usepackage{graphicx}
\usepackage[dvips]{color} 

\definecolor{g-blue}{rgb}{0.83,0.95,1}
\definecolor{Blue}{rgb}{0.5,0.5,1}
\definecolor{DarkBlue}{rgb}{0.00,0.00,0.58}
\definecolor{g-yellow}{rgb}{1,1,0.7}
\definecolor{g-green}{rgb}{0.9,1,0.9}
\definecolor{green}{rgb}{0,0.6,0}
\definecolor{Green}{rgb}{0,0.4,0}
\definecolor{cyan}{rgb}{0,0.7,0.7}
\definecolor{black}{rgb}{0,0,0}
\definecolor{grey}{rgb}{0.4 ,0.4 ,0.4 }

\renewcommand{\sb}[1]{_{\text {#1}}}  
\def\Sb#1{_{\scriptscriptstyle\rm{#1}}}

\newcommand{\C}[1]{{\mathcal{#1}}}    
\def \ed {\end{document}}
\def\Fbox#1{\vskip1ex\hbox to 8.5cm{\hfil\fboxsep0.3cm\fbox{%
  \parbox{8.0cm}{#1}}\hfil}\vskip1ex\noindent}  
\def\<{\left\langle}    \def\>{\right\rangle}
\def\({\left(}          \def\){\right)}
\def \[ {\left [} \def \] {\right ]}
\def\~{\widetilde}
\begin{document}
\title{Evolution of a Neutron-Initiated Micro-Big-Bang in superfluid \3-B}

 \author{ Yury M. Bunkov$^1$, Andrei I. Golov$^2$, Victor S. L'vov$^3$, Anna Pomyalov$^3$ and Itamar Procaccia$^3$}

 \address{$^1$  Universit\'{e} Grenoble Alpes, Institut NEEL, F-38042 Grenoble, France
     \\ $^2$School of Physics and Astronomy, The University of Manchester, Manchester M13 9PL, UK \\ $^3$Department of Chemical Physics, The Weizmann Institute of Science, Rehovot 76100, Israel}

\begin{abstract}
 A nuclear capture reaction of a single neutron by ultra-cold superfluid $^3$He results in a  rapid overheating followed by the expansion and subsequent cooling of the hot subregion, in a certain analogy with the Big Bang of the early Universe. It was shown in a Grenoble experiment that a significant part of the energy released during the nuclear reaction was not converted into heat even after several seconds. It was thought that the missing energy  was stored in a tangle of quantized vortex lines.   This explanation, however, contradicts the expected lifetime of a  bulk  vortex tangle, $10^{-5}-10^{-4}\,$s, which is much shorter than the observed time delay of seconds. In this Letter we propose a scenario that resolves the contradiction: the  vortex tangle, created by the hot spot, emits isolated vortex loops that take with them a significant part of the tangle's energy. These loops quickly reach the container walls. The dilute ensemble of vortex loops attached to the walls can survive for a long time, while the remaining bulk vortex tangle decays  quickly.

\end{abstract}
\maketitle


 \noindent
\emph{\textbf{Introduction.}} It is generally presumed that the Universe started with a Big Bang, expanding subsequently very rapidly while cooling, going through a cascade of gauge phase transitions, in which the four fundamental forces of nature separated out. Today the Universe exhibits inhomogeneous large-scale structures: galaxies that form clusters, arranged in turn in super-clusters such as the ``Great Wall''. The clustering takes the form of long chains or filaments, which are separated  by large voids, regions empty of visible mass. These inhomogeneities might have been nucleated during a rapid non-equilibrium transition known as a ``\emph{quench}". Many types of topological defects  can be created at such transition: domain walls, cosmic strings  and monopoles \cite{Zeld,Kibble}. Cosmic strings have received particular attention among cosmologists since they provide possible seeds for galaxy formation. All these motivate controlled laboratory quench experiments. Unfortunately, attempts of using superfluid $^4$He \cite{4he} failed to observe  any vortices due to various reasons \cite{4he2,Skrbek2006}. In superconductors \cite{Polturak} and in
ferromagnetics~\cite{BEC}  the formation of defects was observed; however, their symmetries   are quite different from that in our early Universe.

Liquid  $^3$He below the temperature $T\sb c \simeq  10^{-3}$K exhibits   a phase transition to superfluidity  characterized by a simultaneous breaking of the orbital, spin and gauge symmetries that are thought to be  the best analogy to those broken after the Big Bang~\cite{VolovikBook}. Moreover, $^3$He can be  locally heated by a nuclear fusion reaction of single low-energy neutron with $^3$He nucleus:  n + $^3$He $\rightarrow$ $^3$H +p.  Each capture deposits the energy of 764\,keV which thermalizes within a
distance on the order of 30\,$\mu$m \cite{LancTeor}
  creating a `hot spot' of normal $^3$He which quickly expands and cools down giving birth to  quantized vortices. This was observed in the Helsinki~\cite{3HeH,3HeHRew} and Grenoble~\cite{3HeG}  experiments in agreement with the predictions of the Zurek modification~\cite{Z}  of the Kibble scenario~\cite{Kibble} of the cosmic strings formation after a Big Bang. Later the analogy between quantum vacuum of $^3$He and Universe was further clarified and summarized in the book by Volovik~\cite{VolovikBook}. In particular, according to Bunkov-Timofeevskaya ``cosmological" scenario~\cite{1998-BT,2000-Bun} the quench in  $^3$He results in   seeds of both A- and B-phases.  Their  initial evolution is determined mainly by the  initial   densities of the seeds.
  Only at a final stage  the difference between  the free-energy densities of A- and B-phases becomes  important~\cite{8}. This   difference  is considered as an  analog of the  hidden Dark Energy of our Universe \cite{VolovikBook,9,10,11}.  Furthermore, after a recent publications of  the new analysis of the results of WMAP and the Plank satellite missions of  cosmic microwave background radiation~\cite{12} this idea becomes popular~\cite{13}.

All these  made the neutron--$^3$He quench experiment a promising physical model of the early evolution of the Universe (or Multiverse?). Therefore a careful analysis of the Grenoble experiment~\cite{3HeG}  in  light of current understanding of the quantized vortex dynamics in superfluid $^3$He is required.  This is the subject of this Letter, which includes  a detailed discussion  of relevant importance of numerous relaxation mechanisms of the tangle of quantized vortex lines in the  $^3$He hot-spot after the quench. In particular we suggest a tangle evolution scenario that resolves a long-standing contradiction between the very short expected lifetime of a  bulk  vortex tangle and the very long lifetime of the hidden (presumably in these vortices) energy. This Letter also formulates  a set of new questions concerning the creation and evolution of a vortex tangle in a neutron--$^3$He quench, that may shed more light on the initial evolution of the Universe after the Big Bang.

\noindent
\emph{\textbf{ The Kibble-Zurek quench scenario  in homogeneous superfluid $^3$He}}.
In   a homogeneous quench, the new phase begins to form simultaneously in  many  independent  regions of the system.  Kibble \cite{Kibble} suggested that inhomogeneity  may originate from fluctuations.  With the growth of coherent regions of the low-temperature phase, they begin to  come in contact with  each others. At the boundaries, where different regions meet, the order parameters do not necessarily match each other, and consequently a domain structure forms. If the broken symmetry is the $U(1)$ gauge symmetry,  these are domains with different phases of the order parameter. Such a random domain structure reduces to a network of linear defects, which are quantized vortex lines in superfluids and superconductors, or cosmic strings in the Early Universe. If the broken symmetry is more complicated, as is the case of $^3$He superfluids, then defects of different dimensionality and structure can form.  Later Zurek proposed~\cite{Z,zurek} a phenomenological approach allowing to estimate the mean intervortex distance
  \begin{equation}\label{tQ}
\ell  \simeq f \xi_0  (\tau\Sb Q/\tau \Sb T)^{1/4}\,, \quad\tau\Sb Q = T\sb c/ [d T(t)/ dt]_{T=T\sb c}\,,
\end{equation}
 and  the resulting vortex-line density $\C L=\ell^{-2}$ (the vortex length per volume).  Here $\xi_0$ is the zero-temperature coherent length, $\tau\Sb T$ is the thermal relaxation time  of the ordered phase  and $\tau\Sb Q $ is the quench time.  The prefactor $f>1$ accounts for the fact that the na\"{\i}ve random walk arguments, leading to Eq.\,\eqref{tQ}  without  $f$, underestimate the density $\C L$.
Early estimates of $f\sim 10$, reported in the past~\cite{47,48,49}, is probably too large. The estimate  $f\simeq  2.5$ was suggested last year~\cite{50}.
Notice that the expected value of $f$ depends on the details of the interatomic potential~\cite{50}, and these are not well known. Thus also this estimate cannot be considered as final.

%
\noindent
\emph{\textbf{The hot-spot evolution in the Grenoble experiment}}.
The Grenoble experiments~\cite{3HeG} were conducted in a cubic box of size $X=5\,$mm   at a background temperature $T_0\simeq 0.1\,T\sb c \simeq  100$\,$\mu$K, at which the superfluid  $^3$He may be considered as a quantum vacuum with an extremely dilute gas of thermal excitations (Bogoliubov quasiparticles). The energy of $E_0=764$ keV deposited  by reaction increases their number.
After a  delay of about $1$\,s their  residual density was measured by a specially designed sensitive bolometer: a   box  of volume  of about 0.1\,cm$^3$ with two vibrating-wire thermometers, immersed in superfluid $^3$He\,\cite{grenoble/thermometry,3Heleshuches}. The detailed analysis of the energy balances  shows that an essential part $\Delta E\sb{st}\equiv E_0-Q\simeq 85\,$keV (at zero pressure) of the   energy $E_0$ was not fully  converted into heat $Q$~\cite{thesis}. It was assumed~\cite{3HeG} that
  the energy deficit $\Delta E\sb{st}$ was stored in the kinetic energy of flow in the form of quantized vortex lines.

  An accurate verification of this hypothesis requires a detailed analysis  of the initial stage of  the spreading dynamics of the temperature following the neutron capture. However it is sufficient for our purposes to do this on a semi-quantitative level by  comparing $E_0$ with the energy of the vortex lines created by the quench.
  To this end we use the continuous media approximation with the temperature diffusion equation
  \begin{equation}\label{dTdt}
 \frac{\partial T(r,t) }{ \partial t} = D\Sb T \left ( \frac{ \partial^2 T}{\partial r^2}  + \frac{2}{r} \frac{\partial T}{\partial r }\right) \ ,
 \end{equation}
with the temperature independent thermal diffusion coefficient $D\Sb T\simeq 5\,$cm$^2$/s\,\cite{archive}. For simplicity we will ignore the possible temperature dependence since are predominantly interested in the later evolution, when the temperature in the center of the hot sphere  is below  $ T\sb c$.
Here we ignore  the fact that the mean free path of thermal excitations is not small compared to the characteristic size of the region in which the products of the neutron absorption thermalize.
The self-similar solution of this equation is:
\begin{equation}
  T(r,t)=T_0+ \frac{E_0}{C_{\rm v}(4\pi   D\Sb T t)^{3/2}}  \exp \Big( -\frac{r^2}{4D\Sb T  t} \Big)\,,
\label{sol}
\end{equation}
where  $C\Sb V = C\Sb V (T_{\rm c}) \simeq  5.83\times 10^3$\,erg\,K$^{-1}$cm$^{-3}$ is the specific heat per unit volume \cite{CV}.  Eq. \eqref{sol} allows us to estimate   the quench time  $\tau\Sb Q\simeq 0.4\, \mu$s  and to see that the temperature in the center of the hot sphere drops below $T_{\rm c}$ in some $t_0=3 \tau\Sb Q/2 \simeq 0.6$\,$\mu$s. It then continues to cool down quickly and reaches 0.5$T_{\rm c}$ in further $\sim 0.4$\,$\mu$s.
The radius of the sphere $  R ( t,  T_*)$, for which $ T( r, t)$ is equal to   $ T\sb c$, depends on $t$ as
\begin{equation}\label{R(t,T)}
R( t,  T_*)=  \sqrt{-  4t  D\Sb T   \ln \big[( T\sb c- T_0)( 4\pi t  D\Sb T )^{3/2} C\Sb V /E_0 \big]}.
\end{equation}
 One can see that $  R( t, T_*)=0$ at $t=0$ and $ t= [E_0/C\sb V (T_*-T_0)]^{2/3}/ 4\pi D\Sb T$.   It reaches its maximum~\cite{FinneJLTP2004}
\begin{equation}\label{Rmax}
R\sb{max} =\sqrt{3/2\pi e}[E_0/(T\sb c-T_0)]^{1/3}
\end{equation}
 at
\begin{equation}\label{tmax}
t\sb{max}= [E_0/C \Sb V (T\sb c-T_0)]^{2/3}/4 \pi e D\Sb T.
\end{equation}
 For $T_0=0.1 T\sb c$  this gives   $ R\sb {max}\approx  26\,\mu$m.

\noindent
\emph{\textbf{Estimate of the initial   vortex-tangle energy $\bm E\sb{\bf vor}$}}.
Taking in  Eq.\,\eqref{tQ} the phase relaxation time  $\tau\Sb T\approx 1.3\,$ns~\cite{Greywall1984}, the quench time $\tau\Sb Q\simeq 0.4\, \mu$s and $f\simeq 2.5$, we estimate the theoretical distance between the vortices, created by the Kibble-Zurek mechanism, as $\ell\simeq 10.8\, \xi_0$. With  the coherent length of  $^3$He $\xi_0  \simeq 0.077 \,\mu$m   this corresponds to $\ell\simeq 0.83 \,\mu$m and to the density of vortex lines
$\C L\simeq 1.5 \times 10^8$cm$^{-2}$. Hence, the total vortex length inside the hot sphere (with $T > T\sb c$) of radius  $R\sb {max}\simeq 26\,\mu$m is  about $L=\C L  \, (4\pi R\sb {max}^3/3) \simeq 11\, $cm.

 The energy of this tangle may be estimated  by assuming   that the vortex orientations are uncorrelated at separations
above $\ell$  (i.\,e. there is no large-scale flow). Then we can use the energy of a quantized vortex line per unit length $\gamma =\gamma_0 \ln (\ell/\xi_0)$. Here $\gamma_0=\rho\sb s \kappa^2/4\pi\simeq 1.76\,$keV/cm (for the superfluid component density $\rho\sb s$ equal to the total $^3$He density, $\rho=0.0814\,$g/cm$^3$) and  the quantum of circulation $\kappa= h /(2\, m_3)=6.6\times 10^{-4}\,$cm$^2$/s (here $m_3$ is the atomic mass of $^3$He). With $\ell \simeq 10.8\,\xi_0$, we arrive at $\gamma\simeq 4.2 \,$ keV/cm that results in the vortex energy $ E\sb{ vor}\simeq 46\,$keV, which is comparable with the experimentally observed residual energy deficit $\Delta E\sb{st}\simeq 85\,$keV.  Baring in mind that Eq.~\eqref{tQ}   gives only an order-of-magnitude estimate~ of $\ell$, we have to consider this quantitative agreement between $ E\sb{vor}$ and $\Delta E\sb{st}$  as a success of the Kibble-Zurek scenario. Now we need to analyze the characteristic times of different channels of dissipation of $ E\sb{vor}$.

 \noindent \emph{\textbf{Decay and diffusion of
vortex tangle}}.  The free evolution of the vortex tangle in the contunious media approximation can be described by the phenomenological Vinen's equation~\cite{vinen}, supplemented by the diffusion term~\cite{vanBeelen1988, Geurst1989, Vinen2003, Nemirovskii2010},
 \begin{equation}
  \frac{\partial  {\C L}(\bm r,t)}{\partial t } = -\nu'\C L^2 + D\Sb L\nabla^2 \C L.
\label{L_Vinen}\end{equation}
Here  $\nu'\simeq 0.1\kappa$ \cite{WalmsleyPRL2008} is the effective kinematic viscosity, and the estimates of the diffusion coefficient $D\Sb L$ vary between $0.1\kappa$ \cite{Vinen2003} and $ 2.2\kappa$ \cite{Nemirovskii2010}. This equation has two characteristic time scales: the decay time of a homogeneous tangle $\tau\sb{dec}\simeq  1/ [\nu' \C L(0)]= \ell_0^2/\nu'$ and   the diffusion time  of a sphere of initial radius $R(0)$, $\tau\sb{dif}\simeq  R(0)^2/D\Sb L$. Having  in mind that $\nu' \sim D\Sb L$, but in the initial tangle  $\ell_0\simeq 0.8 \, \mu$m$ \ll R(0)\simeq 26\,\mu$m, we conclude that
 $\tau\sb{dec}\ll \tau\sb{dif}$. This means that the diffusive spreading  is irrelevant for the problem at hand and the tangle decays in the time $\tau\sb{dec}\simeq   \ell_0^2/\nu'\simeq   10^{-4}$s! In other words, it is impossible to preserve
the initial energy of the tangle, $\simeq 85\,$keV, for longer than $\sim 10^{-4}$ if the tangle is confined to a sphere of radius $R(0)$ as small as $ \sim 26\, \mu$m.

 \noindent \emph{\textbf{   Emission (evaporation) of ballistic vortex loops}}
Numerical simulations\,\cite{Barenghi2002,KN-2012} show that the radial profile of vortex density $\C L(r)$ has a steep drop in an external shell of width $\ell$ near its boundary where the continuous media model~Eq.\,\eqref{L_Vinen} fails. This shell may emit small vortex loops of size $\ell$.   Barenghi and Samuels\,\cite{Barenghi2002} come up with the lifetime of tangles in this process,
\begin{equation}\label{barenghi}\tau\sb{em}\simeq \ell_0^2/\kappa\,,
 \end{equation}
 close to the timescale of the bulk decay $\tau\sb{dec}\simeq \ell_0^2/\nu'$. This means that evaporated loops   can take   a substantial fraction of total energy of the tangle.

 The estimate $\tau\sb{em}\simeq \ell_0^2/\kappa$   is independent of the initial radius of the tangle, $R_0$, which is probably only valid for sparse tangles with $\ell_0 \sim R_0$. Below we employ  a simple  model for the dynamics of evaporation of dense tangles, which gives a $R_0$-dependent lifetime. Namely, we approximate any instantaneous configuration of the  outer layer   as an ensemble of vortex loops of mean radius $\sim \ell$, half of which are moving outwards. For these, the probability of reconnecting with another loop becomes small, and they escape into the open space. For simplicity we assume  a uniform density $ \C L$ (no bulk diffusion), no bulk decay, no counterflow. We approximate vortex loops by rings of radius $\sim \ell$ that travel in all direction with velocity $v \sim \kappa/\ell$ and have bulk mean free path $\sim \ell$. All prefactors of order unity are dropped.

The tangle's outer shell of thickness $\sim \ell$ disappears in time  $\sim \ell^2/\kappa$. About half of the loops escape into open space (their fraction may be enhanced if the vortex loops are outwardly-polarized due to their interaction with the thermal counterflow).
We thus have $d R / dt  = - \kappa \big /  \ell$,
 giving   $R(t)\simeq R_0 -  \kappa \,t/\ell$.  The radius collapses [to $R(\tau\sb {em})=0$] in time
 \begin{equation}
\tau_{\rm em} \simeq  R_0\ell_0/\kappa  \sim 10^{-3}{\rm \,s}\ .
	\label{tau_e}
\end{equation}
This approach gives a result, which is longer than Eq.~\eqref{barenghi} (because $R_0>\ell_0$), yet still much shorter than the experimentally observed time $\sim 1$\,s.

The bottom line is that the evaporation should spare a substantial part of the initial energy in the form of vortex loops of size of order $\ell \sim \C L_0^{-1/2}$ -- which are expected to reach the container walls in some $\tau \sim X\ell /\kappa \sim 2\times 10^{-2}$\,s. This conclusion is supported by recent numerical simulations of vortex reconnections by Kursa {\it et al.} \cite{Kursa2011} and of the decay of a vortex tangle by Kondaurova and Nemirovskii~\cite{KN-2012}, that  show that indeed the leading mechanism of the vortex-line (and, correspondingly, energy) loss in a compact tangle at zero temperature might be the evaporation of vortex loops from its surface.

\noindent \emph{\textbf{ Pinned or frustrated remnant vortices}}
Upon the loop's arrival at the flat container wall at an arbitrary angle, some part of its energy (corresponding to the normal-incidence component of its impulse) might be lost into sound, while the part of energy corresponding to the sideway sliding of the remaining semi-loop should be preserved for long time. Until the surface  density of these loops becomes sufficient to create a developed tangle with frequent reconnections (estimates show that this will never happen), this energy will not be dissipated quickly.

In a container with corners and generally rough walls (which facilitate pinning), the vortices terminated at the wall can become metastable \cite{PLTP2008}.  As the experimentally observed length of metastable vortex lines is independent of pressure \cite{Bunkov2}, we might speculate that this is typical for the particular geometry (size), not pressure. For instance, it was shown by Awschalom and Schwartz \cite{AwschalomSchwartz1984} that in superfluid $^4$He the amount of pinned remnant vortex lines, upon the decay of a larger number, is quite universal, $ \C L_0 \leq 2\ln({X/\xi_0})/X^2\approx 19/X^2$. Even though vortex pinning is much weaker in \3-B, one might still expect these scaling arguments  to work as well.  For the container size $X\sim 5$\,mm,  the total length of remanent vortices can thus be as high as $L \sim \ln({X/\xi_0})X \sim 10$\,cm (that corresponds to stored energy $\gamma_0\ln(X/\xi_0) L \sim 160$\,keV), independent of pressure.

\noindent \emph{\textbf{ Conclusion}}.
We discussed the evolution of a tangle of quantized vortex lines of the initial radius $R_0\sim 26$\,$\mu$m and energy $\sim 85 $\,keV created after a strongly non-equilibrium rapid quench from the normal into the superfluid phase of liquid $^3$He. The tangle disappears within about $10^{-3}$\,s due to two processes of comparable efficiency: bulk decay of quantum turbulence and evaporation of isolated vortex loops away from the tangle. The evaporated vortex loops arrive at the container walls in $\sim 10^{-2}$\,s where they remain in a metastable state for a very long time while keeping a substantial fraction of the initial energy of the vortex tangle. The ``hidden energy'', detected in the Grenoble calorimetric experiments with the time response of $\sim 1$\,s, should thus be comparable with the initial energy of the vortex tangle nucleated upon the quench (Micro-Big-Bang).

The presented  scenario of the Micro-Big-Bang, although  looking feasible,  stresses  significance of many  problems  that require further quantitative analysis.   Among them:
-- accounting for the temperature dependence of the thermal conductivity,  the heat capacity and  the counterflow during the cooling process;   -- estimating
    the critical temperature of the time-dependent, space-inhomogeneous superfluid phase transition;   -- generalization of   the Kibble-Zurek scenario for    more complicated
    symmetries [than the simplest $U(1)$] of quantum vacuum in $^3$He and of the early Universe;    -- calculation of  the relative fractions of the initial vortex length decaying inside  the tangle and evaporating from it;  -- modeling of the collective dynamics of  short vortex loops, pinned to the surface, etc.

     Resolving  these and related problems  may help shedding more light on the intriguing problem of fundamental importance -- the evolution of the early Universe.\\
\noindent \emph{\textbf{Acknowledgements}}.
This work had been supported in part    by  the EU 7$^{th}$ Framework
Programme (FP7/2007-2013, Grant No. 228464
Microkelvin and by  the Minerva
Foundation, Munich, Germany.

\appendix

\subsection*{Appendix A: Energy balance in Grenoble experiment}\label{a:A}
The  energy balance in the Grenoble  experiment has been discussed in \cite{3Heleshuches,thesis}. It was concluded that alternative sources of possible losses can not explain the value of the hidden energy. Here we mention several such processes.

There are defects, other than 1D vortex lines, that could be generated in the superfluid transition: 0D monopoles (boojums) and 2D solitons (domain walls). However, the energy of the former is very small and the formation of the latter can not be foreseen in the geometry of the experiment.

The thermal boundary resistance between the bolometer wall and the superfluid is enormous at these temperatures due to the Kapitza resistance, so any losses via thermal contact are out of question.

The next process can be the ionization of atoms and their scintillation.  We should note that  ionized atoms form  dimers. The spin-singlet dimers radiate the UV radiation while triplets do not.   Experiments performed in liquid $^4$He with electron radiation have shown that helium is quite a good scintillator.   It radiates about  6--8$\%$ of the total energy deposited by high energy electrons in $^4$He, see Ref.~\cite{UV}. One may suggest that for  $^3$He this value should be comparable. But it is not the case.  It is well known, particularly from recent experiments for Dark Matter search, that the ionization after the nuclear recoil is about 3 times smaller than for an electron for the same deposited energy. This principle is used for the separation between the Dark Matter and light particles events, see for example~~\cite{DM}. Taking this circumstances into account we may conclude that the ionisation loses are below 3\%. But we should take also the energy of triplet dimers. The singlet dimers ($25\%$ of states) decay very quickly with the radiation of UV. The rest, the triplet dimers ($75\%$)  live much longer and are quenched on the walls of the cell. The energy of triplet states returns to quasiparticles with a delay of about few seconds~~\cite{Bunkov2}. In conclusion the ionization energy for nuclear recoil events is limited by about $10 \%$


\begin{thebibliography}{99}
\bibitem{CV}
    As the temperature decreases below $T_{\rm c}$, the specific heat  firstly jumps up to about three times te value of the normal phase, $C\Sb V (T)$, and then decreases to about one-third of $C_{\rm v}(T_{\rm c})$ at  $T=0.5T_{\rm c}$.
\end{thebibliography}
\end{document}